\documentclass[graybox]{svmult}

\usepackage{type1cm}        
\usepackage{makeidx}         
\usepackage{graphicx}        
\usepackage{multicol}        
\usepackage[bottom]{footmisc}

\usepackage{newtxtext}       %
\usepackage[varvw]{newtxmath}       

\usepackage{cite}
\usepackage{url} 
\usepackage{subfigure}

\makeindex             

\begin{document}

\title*{Novel non-thermal Ablation Mechanics in the Laser Ablation of Silicon}
\titlerunning{Novel non-thermal Ablation Mechanics in Silicon}
\author{
Dominic Klein, 
Simon Kümmel,
Johannes Roth}
\authorrunning{
D. Klein, 
S. Kümmel,
J. Roth}
\institute{
    Dominic Klein \at Institut f\"ur Funktionelle Materie und
    Quantentechnologien, Universit\"at Stuttgart,\\
    \email{dominic.klein@fmq.uni-stuttgart.de}
    \and
        Simon Kümmel \at Institut f\"ur Funktionelle Materie und
    Quantentechnologien, Universit\"at Stuttgart, \\
    \email{simon.kuemmel@fmq.uni-stuttgart.de}
    \and
    Johannes Roth \at Institut f\"ur Funktionelle Materie und
    Quantentechnologien, Universit\"at Stuttgart,\\
    \email{johannes.roth@fmq.uni-stuttgart.de}
    }

\maketitle

\abstract*{
We investigate the non-thermal material dynamics of strongly excited silicon
during ultra-fast laser ablation. In contrast to metals, silicon shows strongly excitation-dependent interatomic bonding strengths, which gives
rise to a number of unique material dynamics like non-thermal melting, Coulomb
explosions and altered carrier heat conduction due to charge carrier confinement.
In this study, we report novel non-thermal ablation mechanisms in the ultra-fast single shot laser ablation of silicon and 
perform large scale massive multi-parallel simulations on experimentally achievable length scales with atomistic resolution. 
For this, we model the ultra-fast carrier dynamics utilizing the Thermal-Spike-Model coupled to Molecular Dynamics simulations and include
the accompanied excitation-dependent non-thermal bonding strength manipulation by application of the excitation-dependent modified Tersoff Potential.
Further, we present first results on the systematic construction of the excitation-dependent phase diagram of silicon by thermodynamic integration.
} 

\abstract{
We investigate the non-thermal material dynamics of strongly excited silicon 
during ultra-fast laser ablation. In contrast to metals, silicon shows strongly excitation-dependent interatomic bonding strengths, which gives
rise to a number of unique material dynamics like non-thermal melting, Coulomb
explosions and altered carrier heat conduction due to charge carrier confinement.
In this study, we report novel non-thermal ablation mechanisms in the ultra-fast single shot laser ablation of silicon and 
perform large scale massive multi-parallel simulations on experimentally achievable length scales with atomistic resolution. 
For this, we model the ultra-fast carrier dynamics utilizing the Thermal-Spike-Model coupled to Molecular Dynamics simulations and include
the accompanied excitation-dependent non-thermal bonding strength manipulation by application of the excitation-dependent modified Tersoff Potential.
Further, we present first results on the systematic construction of the excitation-dependent phase diagram of silicon by thermodynamic integration.
}   

\section{Introduction}
\label{sec:intro} 
The non-equilibrium phenomena in highly excited crystalline solids induced by
strong laser radiation fields have received much attention in recent
years. Despite many theoretical and computational investigations, these
ultra-fast processes are still not well understood, especially in the case of
covalently bonded semiconductors.
Recent publications have reported that covalently bonded semiconductors show bond weakening effects under extreme laser irradiation
and can even be driven into a state of purely repulsive interatomic bonds. 
Such ultra-fast ultra-strong excitation gives rise to a lattice dissolution within the
order of hundreds of femtoseconds, accompanied with highly unstable states of matter \cite{Medvedev15,Zier15,Kiselev17}.
In the context of laser ablation, these non-thermal effects give rise to a complex interplay of competing ultra-fast thermal and non-thermal material dynamics, 
altering the dominating ablation mechanisms and composition of the ejected material. 
In this work, we investigate the implications of non-thermal effects on the material dynamics during laser ablation of silicon in two ways. 
First, we perform massive multi-parallel continuum-atomistic simulations.
Here, we apply the so-called Thermal-Spike-Model (TSM) \cite{klein2021} to model the ultra-fast electronic excitation and transport effects within 
the carrier subsystem and depict the long-time macroscopic material dynamic by the Molecular Dynamics (MD) approach.
For the inclusion of non-thermal effects, we implement the electronic excitation-dependent modified Tersoff Potential (MOD*) \cite{Kiselev17,Kiselev2016}.
Within this simulation framework, we perform ultra-fast single-shot laser ablation simulations on a bulk silicon samples under irradiation of a spatial and temporal Gaussian pulse profile and classify the occurring novel non-thermal ablation mechanisms. 
Further, we present first investigations on the phase-diagram of strongly excited silicon by thermodynamic integration \cite{Kirkwood_1935_TI, Menon_2021_calphy}. Doing so enables us to get more insight into the interplay between the degree of excitation and the structural rearrangements in the material. 

\section{Simulation Setup}

\begin{figure}[htb]
\centering
\includegraphics[scale=.99]{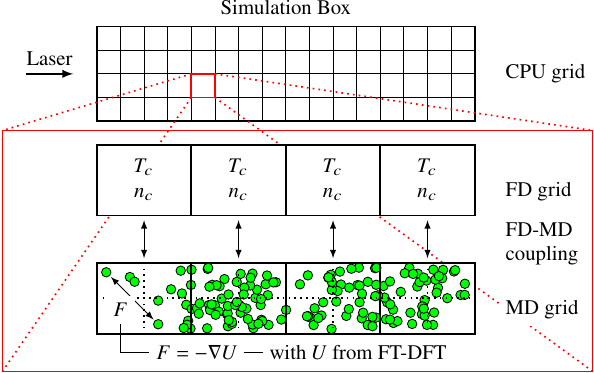}
%
%
\caption{Schematic presentation of simulation composition in two dimensions. In practice, all grids are three-dimensional. Implementation of periodic boundary conditions and communication schemes are not shown.}
\label{fig:setup}
\end{figure}

In this work, we model the process of laser ablation on silicon using our in-house simulation package  \texttt{IMD} \cite{Roth2019}.
We solve the heat transport equations for the charge carrier density in
semiconductors, as well as the charge carrier transport equation by
a continuum Finite-Difference (FD) method coupled to MD
simulations using the TSM implemented in \texttt{IMD}.
A detailed description of the method and the applied model can be found in \cite{kleindiss1,klein2021}. 
A simplified sketch of the simulation setup is shown in Fig.~\ref{fig:setup}.
The simulation box is spatially divided into CPU cells, distributed over the number of physically available CPUs. Each CPU cell contains the local information of the charge carrier subsystem. On the FD grid, observables of the electronic subsystem, like local carrier temperature $T_c$ or carrier density $n_c$ are calculated within the TSM \cite{kleindiss1,klein2021}. For simplicity, only two- or one-dimensional grids are sketched here, while in practice all grids are three dimensional. Each FD cell is further divided into MD cells, which contain the positions and velocities of their associated atoms. The FD and MD subsystems exchange energy by electron-phonon coupling and lead to subsequent heating of the atoms in the embedded MD cell.
The force acting on each atom is modeled 
using the classical modified Tersoff Potential (MOD) \cite{Kumagai2007} and the excitation dependent modified Tersoff Potential (MOD*) \cite{Kiselev2016,Kiselev17}.
The latter is parametrized from Finite-temperature (FT) Density Functional Theory (DFT) calculations to include non-thermal effects.
We choose a spatial and temporal Gaussian laser beam with the surface intensity 
\begin{align}
 I(t,r) =  (1-R) &\underbrace{\sigma \cdot \frac{4 \ln 2}{\pi b^2}   \exp \left( -4 \ln 2 \frac{r^2}{b^{2}}  \right)}_{\equiv \sigma(r)} \vphantom{\sqrt{\frac{4 \ln 2}{\pi t_p^2}}} \nonumber \\
 &\cdot \sqrt{\frac{4 \ln 2}{\pi \tau^2}} \exp \left( -4 \ln 2 \frac{(t - t_0)^2}{\tau^{2}}  \right) \vphantom{\sigma \cdot \frac{4 \ln 2}{\pi b^2}   \exp \left( -4 \ln 2 \frac{r^2}{b^{2}}  \right) \vphantom{\sqrt{\frac{4 \ln 2}{\pi t_p^2}}}}  \label{pulse}
\end{align}
at time $t$, a spatial full width at half maximum of $b = 500$ nm, a temporal full width at half maximum of $\tau = 100$ fs and a fluence of $\sigma$. 
$R$ is the surface reflectivity and $t_0$ the time of peak intensity. The optical parametrizations are chosen to reflect a wavelength of $\lambda = 800$ nm
and are summarized in \cite{kleindiss1,klein2021}. \\

To create the phase diagrams, we use the implementation of the thermodynamic integration in   \texttt{calphy} 1.0.0 \cite{Menon_2021_calphy} which in our case uses the stable version from 29 September, 2021 of  \texttt{LAMMPS} \cite{Thompson_2022_LAMMPS} to perform the MD part of the thermodynamic integration.
For a given pressure and atomic structure, the atomic system is relaxed in an MD simulation, then the temperature is changed over time. From the work done, the temperature-dependent Gibbs free energy can be calculated. The intersection of the Gibbs free energy curves of two structures indicates the phase transition between the two. Repeating this for several different pressures and all pairs of phases, we end up with a complete phase diagram.

\section{Results}

In Sect.~\ref{sect:ablech}, we present the novel ablation mechanisms obtained by laser ablation simulations and in 
Sect.~\ref{sect:phase} we present first data sets on the systematic construction of the phase diagram of strongly excited silicon by thermodynamic integration.

\subsection{Novel Ablation Mechanics and Ablation Depth} \label{sect:ablech}
\begin{figure}[htb]
\centering
\subfigure[MOD]{\resizebox{0.49\textwidth}{!}{ \includegraphics[scale=1.0]{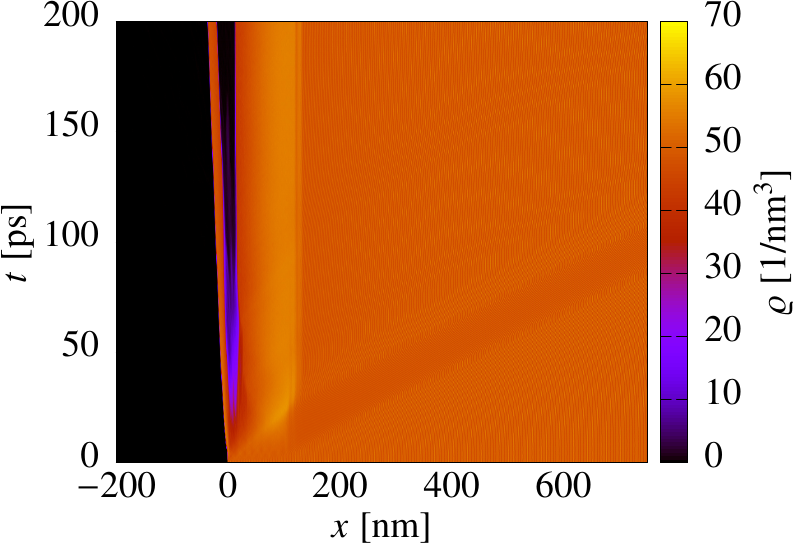}}}
\subfigure[MOD*]{\resizebox{0.49\textwidth}{!}{ \includegraphics[scale=1.0]{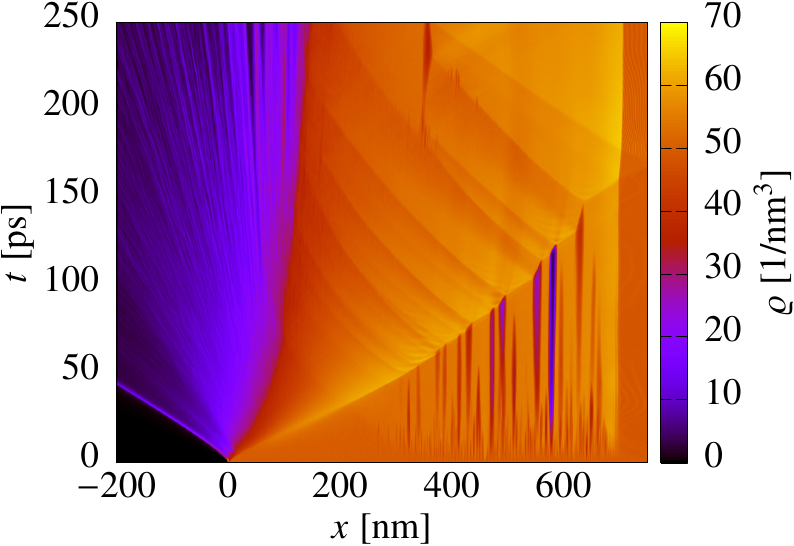}}}
\caption{Comparison of time resolved density histograms for $\sigma = 0.672$ J/cm$^2$ for the MOD and MOD* interaction potential.}
\label{fig:mecha}
\end{figure}

We perform laser ablation simulations on bulk silicon sample Si1D (Sect.~\ref{sect:res}) applying the established MOD and the excitation-dependent MOD* interaction potential separately. Si1D shows a lateral extension of $22$ nm and thus can be seen as a quasi-1D simulation geometry set in the center of the ablation crater.
The resulting time resolved density histograms in laser direction $x$ are presented in Fig.~\ref{fig:mecha}. The displayed laser fluence $\sigma = 0.627$ J/cm$^2$ corresponds to roughly 3.5 times the ablation threshold.

The usage of the MOD potential depicts the ablation dynamics being dominated by phase explosion, which is a typical ablation mechanism for metals where non-thermal effects play a minor role in material dynamics \cite{rethf17,Balli13}.
The occurring material responses include a shock wave propagating into the sample by the local speed of sound, a heterogeneous melting front propagating into the sample and the formation of a sub-surface overcritical fluid. The overcritical fluid expands into a mixture of droplets and vapor, ejecting a the surface layer, hence the name phase explosion.

The inclusion of bond weakening non-thermal excitation effects by applying the MOD* potential, however, predict novel ablation mechanisms. 
We classify the novel ablation mechanics as follows
\begin{itemize}
 \item Non-thermal evaporation: Non-thermal evaporation is an extreme version of non-thermal melting, where anti-bonding states dominate the interatomic energy landscape. A sufficiently high laser excitation first leads to bond weakening and increase in thermal energy. Like in purely classical theories, the material evaporates, when the thermal energy exceeds the potential energy. However, in the case of non-thermal surface evaporation, the evaporation takes place on the sub-picosecond timescale, leading to an instantaneous ejection of vapor.
 \item Pre-shockwave non-thermal melting: In laser ablation, we expect the sample surface to rapidly expand due to material heating, thus generating a pressure wave propagating into the sample. With sufficient heating, this results in a heterogeneous melting front. Classically, the melting front propagation speed is restricted to the local speed of sound. However, we observe a non-thermal melting front, propagating with the local speed of light up to the depth, at which internal fluence still meets the conditions for non-thermal melting.
 \item Non-thermal void formation: Non-thermal void formation is a direct consequence of non-thermal melting. During non-thermal melting, the diamond structure of silicon collapses into a dense amorphous state which is accompanied by an instantaneous drop in local pressure.
 The melting speed $v_\text{melt}= x_\text{melt}/\tau_\text{melt}$ exceeds the reported equilibrium speed of sound $c_s(300$ K$) = 8433$ m/s by three orders of magnitude. As a result, the local material flow is too slow to compensate the resulting vacuum and the material relaxes by the formation of voids. 
\item Void-induced liquid spallation: Void-induced liquid spallation is another consequence of non-thermal void formation. 
In the classical sense, spallation is the effect of reflected shockwaves propagating back to the sample surface, interferes with itself upon surface reflection and causes crystalline or liquid material to be ejected.
In this new mechanism, the non-thermal voids generate new open surfaces on which the pressure wave gets partially reflected.
Additionally, due to the preceding pre-shockwave melting, the material in the laser-affected zone is present in liquid or gaseous form, leading to 
the emission of liquid droplets.
\end{itemize}

\begin{figure}[htb]
\centering
\includegraphics[scale=0.8]{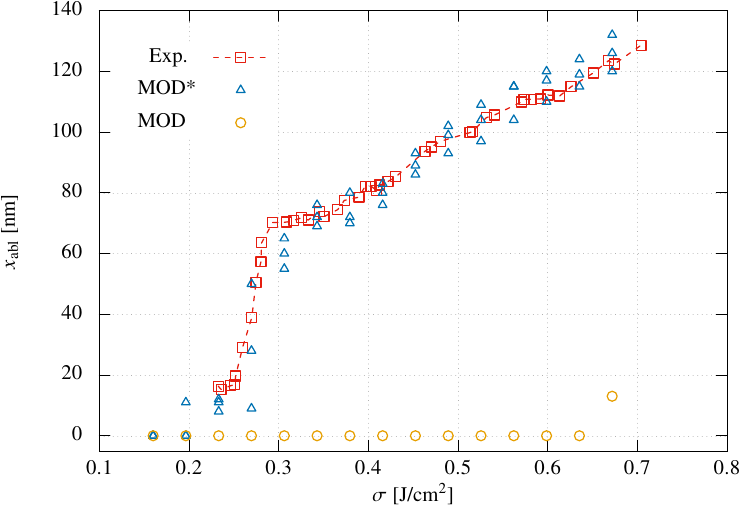}
\caption{Ablation depth $x_\text{abl}$ depending on laser fluence $\sigma$. Blue triangles represent data obtained under the MOD* potential and yellow circles represent the data obtained under the MOD potential. Red squares are experimentally measured ablation depths by Zhang~\cite{Zhang2013}. }
\label{fig:zhang}
\end{figure}

Despite the drastic deviation from currently published ablation mechanisms in laser irradiated silicon \cite{rethf17,Balli13}, the model is not only in agreement with MD-DFT simulations on non-thermal melting times and non-thermal melting thresholds but also with its prediction of ablation depth.
In Fig.~\ref{fig:zhang}, we show the resulting ablation depth applying the MOD and MOD* in the range of $0.072$ J/cm$^2 < \sigma < 0.627$ J/cm$^2$,
as well as experimental data obtained by single shot ablation experiments under identical laser parameters.
The presented data does not only show the significantly better agreement of experimental data and simulation predictions of the MOD* when compared to the MOD but also a nearly 1:1 match to the experimental data underlining the importance of the inclusion of non-thermal effects and validating the novel ablation mechanisms 
introduced above. 

\subsection{Realistic Sample Sizes and composed Simulations} \label{sect:phase}

\begin{figure}[!htb]
\centering
\subfigure[D: $\sigma = 0.26\text{ J/cm$^2$}$\label{simres:2D:abldepth:MATRIX::b}]{\resizebox{0.49\textwidth}{!}{ \includegraphics[scale=1.0]{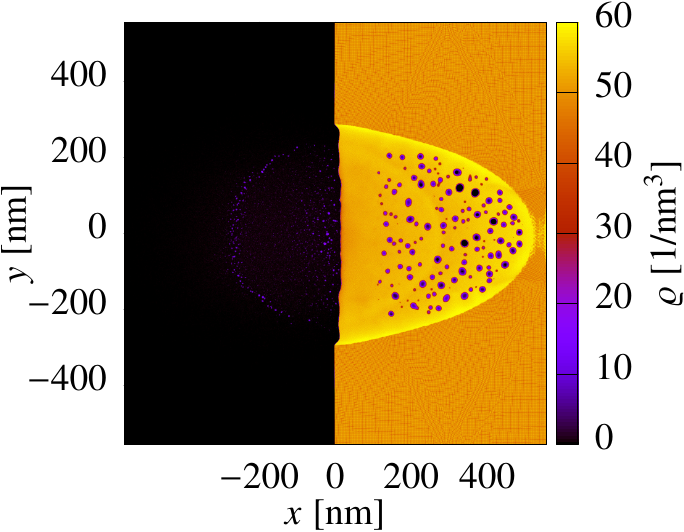}}}
\subfigure[C: $\sigma = 0.26\text{ J/cm$^2$}$\label{simres:2D:abldepth:MATRIX::b}]{\resizebox{0.49\textwidth}{!}{ \includegraphics[scale=1.0]{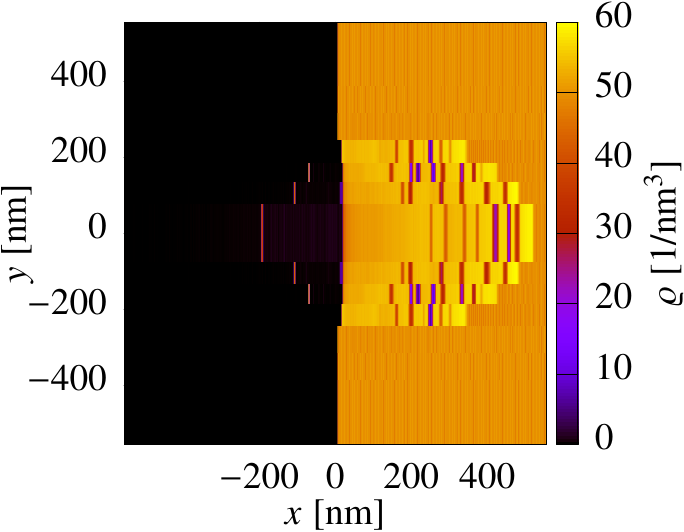}}}
\subfigure[D: $\sigma = 0.42\text{ J/cm$^2$}$\label{simres:2D:abldepth:MATRIX::f}]{\resizebox{0.49\textwidth}{!}{ \includegraphics[scale=1.0]{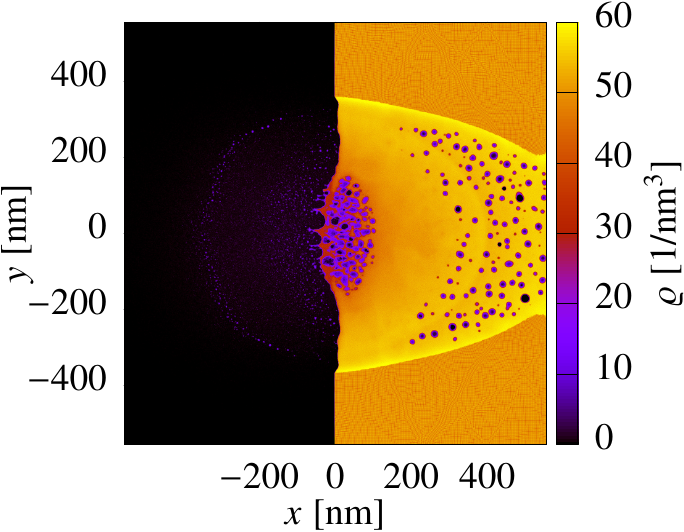}}}
\subfigure[C: $\sigma = 0.42\text{ J/cm$^2$}$\label{simres:2D:abldepth:MATRIX::f}]{\resizebox{0.49\textwidth}{!}{ \includegraphics[scale=1.0]{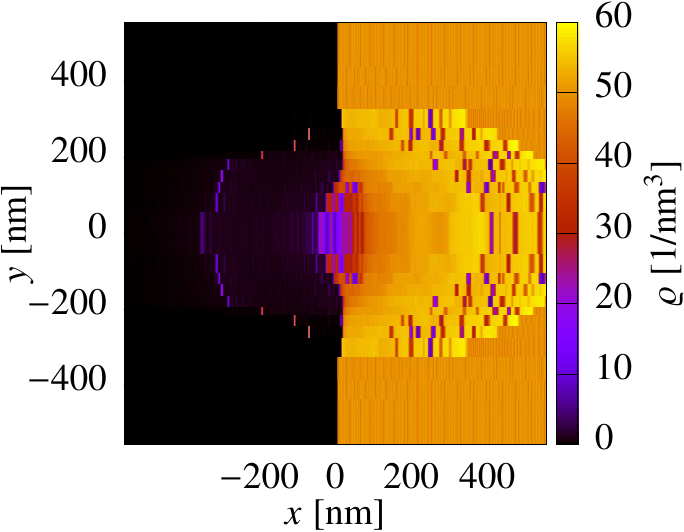}}}
\subfigure[D: $\sigma = 0.67\text{ J/cm$^2$}$\label{simres:2D:abldepth:MATRIX2::f}]{\resizebox{0.49\textwidth}{!}{ \includegraphics[scale=1.0]{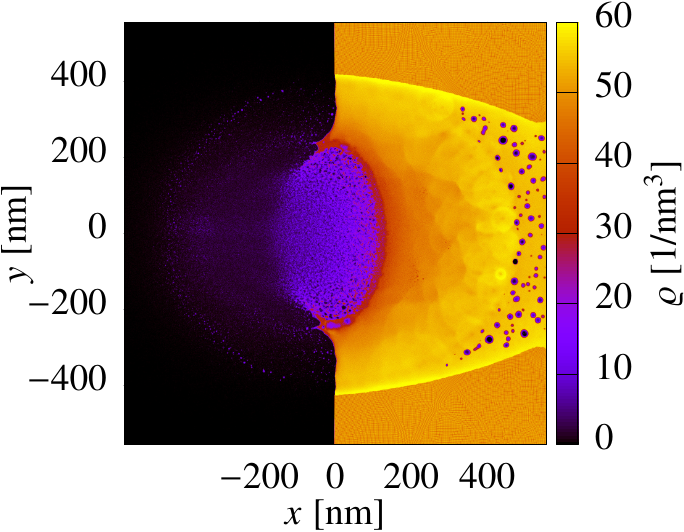}}}
\subfigure[C: $\sigma = 0.67\text{ J/cm$^2$}$\label{simres:2D:abldepth:MATRIX2::f}]{\resizebox{0.49\textwidth}{!}{ \includegraphics[scale=1.0]{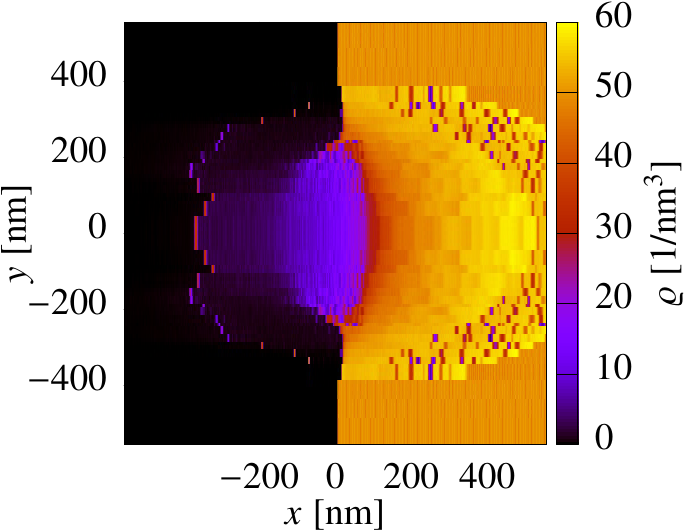}}}
\caption{Snapshots of the spatial density histograms at a simulation time of $t= 100$ ps after peak laser intensity using the MOD* as direct simulations (left D)
on sample Si2D and composed from simulations on sample Si1D (right C) for varing laser fluences $\sigma$.   }
\label{2dcrater}
\end{figure}

The 2020 upgrade from Hazel Hen to Hawk at the High Performance Computing Center Stuttgart and the accompanied substantial increase in performance 
allow for the atomistic simulation of laser ablation on experimentally achievable laser spot sizes. To our knowledge, this was not possible before and we are the first group to perform atomistic laser ablation simulations at length scales that include the complete crater profile. 
In Fig.~\ref{2dcrater}, we present snapshots of the spatial density distribution of laser ablation simulations on sample Si2D (Sect.~\ref{sect:res}) with a lateral extension of over a micrometer, irradiated with a full width at half maximum laser spot size of $b = 0.5$ $\mu$m at a wavelength of $\lambda =800$ nm under usage of the MOD* potential for selected laser fluences at $t= 100$ ps simulation time.
Additionally, we show spatial density distribution snapshots obtained by composition of simulation data on the Si1D sample from the previous section. Here, each density distribution along the $x$-axis corresponds to quasi-1D simulations at peak fluence $\sigma = \sigma(r)$ (see eq. \eqref{pulse}) at identical simulation time. \\ 
Within this simulation framework we observe three notable features:
\begin{itemize}
 \item Silicon shows three distinct ablation mechanisms. For fluences up to $\sigma = 0.35$ J/cm$^2$, the ablation mechanism is dominated by conventional thermal ablation, but the morphology of the remaining crater is dominated by the non-thermal melting pool. 
 For fluences of roughly $0.35$ J/cm$^2 < \sigma < 0.45$ J/cm$^2$, phase explosion becomes the dominating ablation process until it is surpassed 
 non-thermal surface evaporation for fluences $\sigma > 0.45$ J/cm$^2$.
 \item Pre-shockwave non-thermal melting occurs at all laser fluences surpassing the non-thermal melting threshold. For regions with $\sigma(r)$ being below the ablation threshold, this non-thermal melting process leads to the formation of an amorphous rim around the ablation crater. 
 The formation of such amorphous structures is currently reported and discussed in literature \cite{bonse02,werne19}. Despite the efforts, a conclusive explanation of the underlying mechanisms is still missing. However, our simulation data strongly suggests that the occurrence of the amorphous ring structure is a direct result of non-thermal pre-shockwave melting.
 \item The accuracy of composed simulations has often been argued and proposed, but not proven since quasi-2D simulations were not feasible in the past. 
 Our composed laser ablation simulations show remarkable accordance depicting the material dynamics of simulations. 
 We observe that the composed simulations yield identical results in terms of crater and non-thermal rim diameters, as well as ablation depth and placement of the ablation crown. 
 However, the lateral material dynamics features are artificially suppressed in the case of composed simulation.
Such effects are needed to depict effects, such as nano-bump formation or redeposition
of larger clusters slung back by the ablation crown or lateral shockwaves closing non-thermal voids. For fluences above the non-thermal melting threshold, the surface tension of the molt and the ejected droplets become neglectable. This is due to non-thermal evaporation being the dominating ablation mechanic. Composed simulations yield comparable results to the direct simulation setup when only the ejected material and macroscopic crater features are of interest.

This is especially noteworthy since laser ablation simulation composition shows to be a possibility to save enormous amounts of computational resources. 
For the shown composed simulations, we used production run simulations at 17 different laser fluences, accumulating to 83,814.13 core-h of computation time. 
From the obtained data set, all fluences below the chosen peak fluence of $\sigma_p = 0.69$ J/cm$^2$ can be composed in arbitrary pulse shapes. 
In comparison, a production run for one fluence on sample Si2D consumes 393.926,42 core-h of computation time for an identical simulation time of $t = 100$ ps. 
This translates to roughly $\approx 4.7$ times the computational resources needed as for the complete Si1D data set.
\end{itemize}
A more comprehensive and detailed analysis of the effects discussed above can be found in \cite{kleindiss1}.

\subsection{The Phase Diagram of strongly excited Silicon}
\begin{figure}[!htb]
	\centering
	\includegraphics[width=0.79 \linewidth]{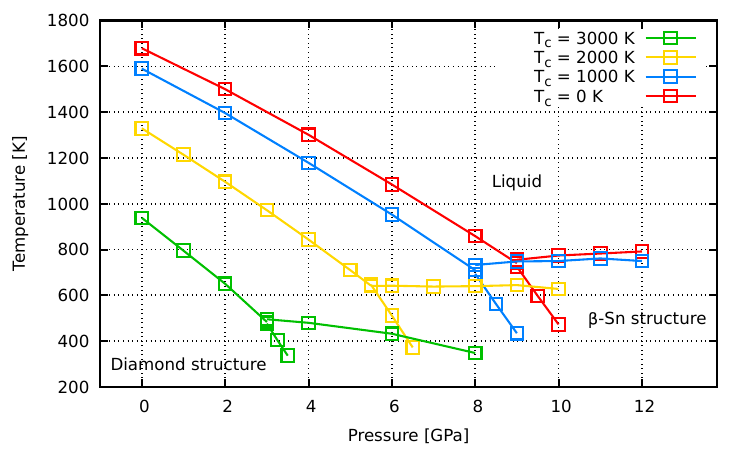}
	\caption{Temperature-pressure phase diagrams of silicon at different electron temperatures.}
	\label{fig:Phase_diagrams}
\end{figure}

To get more insights in the non-thermal effects on the structure, we want to calculate the phase diagrams of silicon at different electron temperatures. This way, we can link the degree of excitation to the structural changes and especially see how the non-thermal melting changes with the degree of excitation. \\
We started by calculating the melting temperature at zero pressure using the standard MOD \cite{Kumagai2007}. With appropriate parameters and a system containing 4096 atoms, we calculate the melting temperature of the diamond structure three times. We find an average melting temperature of 1688.6(12) K which is in excellent agreement with the experimental value of 1687 K \cite{Yamaguchi_2002_Melting_temp}. From this first test, we conclude that a system of a few thousand atoms is large enough and decided to use a system containing 4096 atoms in the diamond structure and the liquid phase, as well as 4000 atoms in the $\beta$-Sn structure, a high-pressure phase of silicon, to keep the accuracy of the statistics approximately the same for the production runs. When repeated, the melting temperature fluctuates by the order of less than 0.1\%. \\
Since we want to create phase diagrams of silicon at several different electron temperatures, we switch from the standard MOD to the electron temperature-dependent MOD* \cite{Kiselev17}. These pressure-temperature phase diagrams at different electron temperatures are shown in Fig.~\ref{fig:Phase_diagrams}. The phase transitions are shifted to lower temperatures at higher electron temperatures. This effect is known as bond weakening and is the result of excitation of electrons into antibonding states which are less stable compared to the bonding states. If the electron temperature in silicon is sufficiently high, it undergoes a phase transition to the liquid phase which is a prominent example of a non-thermal phase transition since the lattice temperature might be nearly unchanged for a few femtoseconds \cite{Kiselev17} after laser-irradiation. This bond weakening can also be seen at high pressures where the $\beta$-Sn structure is stable. \\
We found that the diamond structure is exponentially unstable with the degree of excitation. This also means that non-thermal effect like non-thermal melting become increasingly dominant at high excitations. Unfortunately, it is not possible to find phase transitions into the gaseous state using thermodynamic integration which would have been interesting in the context of laser ablation.

\section{Conclusion}

In this work, we presented novel non-thermal ablation mechanisms predicted by MD simulations under inclusion of excitation-dependent interatomic interactions. 
These novel ablation mechanisms show strong implications on the material dynamics of silicon during the ablation process and suggest that 
silicon shows three distinct ablation regimes dominated by non-thermal melting, phase explosion or non-thermal evaporation. 
By comparison with experimental data, we have shown that the TSM correctly predicts ablation depth and threshold and thus is a huge improvement on currently existing models.
From the electron temperature-dependent phase diagrams, we found the direct connection between the degree of excitation and the instability of the diamond structure in silicon and even the non-thermal melting at sufficiently high excitation. The diamond structure was found to be exponentially unstable with the level of excitement, indicating that non-thermal effects become increasingly dominant in the material dynamics at high excitations.

\section{Computational Resources}
\renewcommand{\arraystretch}{1.6}
\label{sect:res}

In Sect.~\ref{sec:hlsdata}, we present technical details on the performance of laser ablation simulations using \texttt{IMD}. 
In Sect.~\ref{sec:phaseres}, we present performance reviews of the simulation packages used for the thermodynamic integration of strongly excited silicon. 
In Sect.~\ref{consumed}, we summarize the consumed computational resources at the High Performance Computing Center Stuttgart. 

\subsection{Performance of Laser Ablation Simulations using IMD}
\label{sec:hlsdata}
In the current reporting period, we have successfully carried out molecular
dynamics simulations of laser ablation on the silicon samples 
\begin{itemize}
 \item Si1D which shows a size of 2225 nm $\times \, 22$ nm $\times \, \phantom{11}22$ nm and
 \item Si2D which shows a size of \phantom{5}556 nm $\times \, 17$ nm $\times \, 1113$ nm
\end{itemize}
using our in-house simulation package  \texttt{IMD}. The numbers of atoms, applied processors and achieved performance for a laser ablation production run are shown in
Tab.~ \ref{tpg} for each sample. 
\begin{table}[!htb]
  \centering
  \caption{Typical resources applied and performance achieved in CPUseconds 
    (cpus per atom $n$ per simulation-step $\Delta t$).}
  \label{tpg}
\label{tab_scaling}
  \begin{tabular}{l c r r r r } \hline
    simulation & wall time & \#\,atoms  $n$ & \#\,nodes & \#\,processors & performance $[\text{CPUs}/n\Delta t]$ \\ \hline 
    Si1D & 24h & 18,121,592 & 64 & 8,192 & $5.344767 \cdot 10^{-5}$ \\
    Si2D & 24h & 536,870,912& 1024 &  65,536 & $  5.174720 \cdot 10^{-4}$ \\ \hline
  \end{tabular}
\end{table}

\begin{figure}[htb]%
        \centering%
        \includegraphics[scale=0.9]{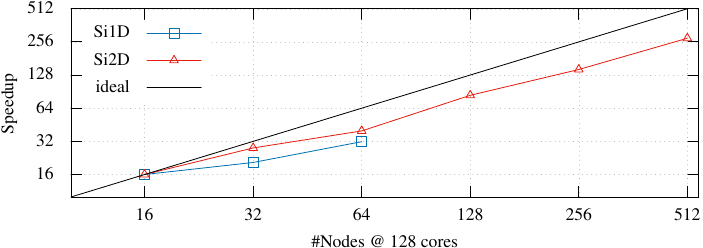}
         \caption{\label{fig_speedup}Scaling behavior of IMD on Hawk at
           HLRS. This data was obtained with a problem sizes given in Tab.~
           \ref{tab_scaling}.} 
        \label{pic:latex-sp}%
\end{figure}

The used in-house simulation package  \texttt{IMD} is a typical molecular dynamics simulation code. As such, the main
computation load is generated in a single main loop where the classical
interactions between atoms are calculated. The program achieves
parallelization degrees of the order of 90 and more percent, depending on the
simulated experimental setup.  The main loop can be fully parallelized with  \texttt{MPI} due to its short range nature
and distributed to any number of processors. The resulting parallelization speedup is only hindered by the communication overhead. 
The scaling behavior for both samples treated with  \texttt{IMD} is shown in Fig.~\ref{fig_speedup} and Tab.~\ref{tab_speedup}. 

\begin{table}[!htb]%
\caption{\label{tab_speedup} Scaling behavior in terms of performance (P) and
  speedup (S) of IMD on Hawk at HLRS. This test was performed with the noted
  number of particles.} 
\hspace*{-1cm}
\setlength{\tabcolsep}{4pt}
\renewcommand{\arraystretch}{1.2}
\centering
\begin{tabular}{c cc cc} \hline
\# Nodes @ & \multicolumn{2}{c}{Si1D} &\multicolumn{2}{c}{Si2D} \\ 
 128 cores & P $[\text{CPUs}/n\Delta t]$ & S & P $[\text{CPUs}/n\Delta t]$ & S\\ \hline  
 16 & $2.688\cdot 10^{-5}$ & 16.0 & $2.998\cdot 10^{-5}$ & 16.0\\
 32 & $3.459\cdot 10^{-5}$ & 20.6 & $5.228\cdot 10^{-5}$ & 27.9\\
 64 & $5.345\cdot 10^{-5}$ & 31.8 &  $7.463\cdot 10^{-5}$ & 39.8 \\
 128 & - & - & $1.570\cdot 10^{-4}$ & 83.8  \\
 256 & - & - & $2.694\cdot 10^{-4}$ & 143.8 \\
 512 & - & - & $5.175\cdot 10^{-4}$& 276.2\\ \hline 
\end{tabular}
\end{table}

\subsection{Performance of the construction of Phase Diagrams} \label{sec:phaseres}

\begin{figure}[!htb]
	\centering
	\includegraphics[width=0.8 \linewidth]{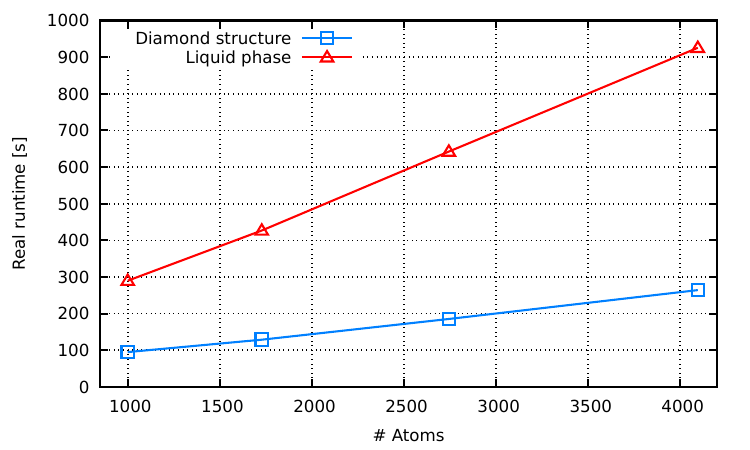}
	\caption{Scaling of calphy with the number of atoms in a diamond structure and the liquid phase.}
	\label{fig:Runtime}
\end{figure}

From calculations with different number of atoms, we find the scaling of the code. As can be seen in Fig.~\ref{fig:Runtime}, the calculations scale linearly with the number of atoms which can be attributed to the linear scaling of all molecular dynamics simulations with  \texttt{LAMMPS} \cite{Thompson_2022_LAMMPS}.

\begin{figure}[!htb]
	\centering
	\includegraphics[width=0.8 \linewidth]{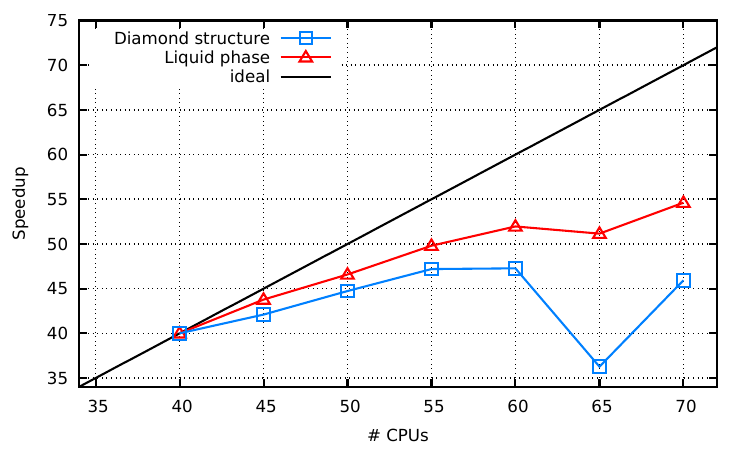}
	\caption{Speedup of the calculations of the melting temperature of silicon at zero-pressure for different number of cores.}
	\label{fig:speedup}
\end{figure}

Since calphy uses  \texttt{LAMMPS} for the molecular dynamics part of the thermodynamic integration which is also the computationally most demanding part, it should be run on several cores. Conveniently, calphy already uses the parallelization schemes implemented in  \texttt{LAMMPS}. Since we're dealing with comparably small systems in the molecular dynamics context, single-node jobs are an obvious choice. \\
In order to find the ideal number of cores to use during the production runs, we repeat the calculation of the melting temperature of silicon at zero-pressure with different numbers of cores. The resulting runtime at different number of cores is shown in Fig.~\ref{fig:speedup}.

As can be seen in Fig.~\ref{fig:speedup}, the best performance is achieved when less than half of a node with 128 cores is used for a single calculation. This allows us to run two calculations at the same time on a single node. A primitive approach would be to always start calculations in pairs. If one calculation is faster than the other, it then has to wait for the slower one to finish before the next pair of calculations is started. Since the runtime of the calculations might vary a lot, we start two calculations simultaneously and track the process IDs of the calculations regularly. This way, if one calculation has finished, we can start a new calculation immediately, ensuring again that only two are running at the same time. \\
To test this, we calculate the melting temperature of a silicon system containing 4096 atoms three times within on job, two calculations at any given time and do this procedure three times in separate jobs for a good statistics of the runtimes. Each calculation was done on 55 cores, using a total of 110  \texttt{MPI} processes per job since we found not using almost all cores of a node to be performing better compared to calculations using all cores. Doing so, we find that each melting temperature calculation now takes an average 427.7(389) s, compared to an average 736.7(291) s when the calculations were done in serial which yields a parallel efficiency of around 86.1\% since we can now run two calculation simultaneously.

\subsection{Consumed Computing Resources} \label{consumed}

Since April 2022, a total of 98,248.26 node-hours have been spent on HPE Apollo
(Hawk). The used resources are summarized in Tab.~\ref{consumedrescources}.
The largest share has been used for the PhD thesis of
Dominic Klein \cite{kleindiss1} studying laser ablation of silicon.
Within the course of his master's thesis, Simon Kümmel used 170.86 node-hours for the study of the phase diagram of highly excited silicon. 
Both studies are subject to the results shown in this work.
Azad Gorgis used 22,984.26 node-hours for his master's thesis on the topic of simulations of the 3D printing of metals and the preparation 
for the project 3DLP-Master (ACID: 44236). 
Dennis Rapp used 667.70 node-h for test runs and optimization of  \texttt{IMD}.

\begin{table}[!htb]
\centering
\begin{tabular}{c r r c}  \hline
User & Resources [node-h] & Rescources [core-h] & Usage \\ \hline 
Dominic Klein & 74,425.44   & 9,526,456.32& Laser ablation \\
Simon Kümmel  & 170.86     & 21,870.08 & Phase Diagrams\\
Azad Gorgis   & 22,984.26   & 2,941,985.28  & Preparation for 3DLP \\
Dennis Rapp   & 667.70     & 85,465.60 & Test runs with  \texttt{IMD} \\
Total         & 98,248.26 & 12,575,777.28 & \\ \hline 
\end{tabular}
\caption{Consumed computational resources for the project LASMD (ACID:1283) in the time period of April 2022 to May 2023.}
\label{consumedrescources}
\end{table}


\end{document}